\documentclass{article}
\usepackage{spconf,amsmath,graphicx,booktabs,multirow,hyperref}
\usepackage{colortbl}

\title{PitchIn: Natural-Language Vocal Coaching from Unstructured Community Feedback}
\name{Hayeon Bang$^{1}$, Hounsu Kim$^{1}$, Joonyong Park$^{2}$, Jung-Jae Yu$^{3}$, Juhan Nam$^{1}$}
\address{$^{1}$KAIST, Republic of Korea \qquad $^{2}$The University of Tokyo, Japan \qquad $^{3}$ETRI, Republic of Korea}

\begin{document}
\ninept
\maketitle
\begin{abstract}
Effective vocal coaching connects performance issues with corrective advice, but recordings paired with expert diagnoses and advice are scarce.
Online singing communities provide a potential source of supervision, but observations and advice are scattered across unstructured comments.
We introduce \textbf{PitchIn}, audio--language models for natural-language vocal coaching, and the \textbf{PeerCoach Dataset}: 14,445 instruction examples covering Focused QA, Overall Feedback, and Recognition QA.
We compare target combinations on Qwen2.5-Omni and train PitchIn on Focused QA and Overall Feedback using Qwen2.5-Omni and MOSS-Music.
On VocalCoachBench, the proportion of requested diagnosis--correction pairs matching an expert issue with valid advice rises from 21.3\% to 61.0\% for Qwen2.5-Omni and from 31.4\% to 51.2\% for MOSS-Music.
Direct training on source comments yields only 19.8\% for Qwen2.5-Omni.
For Qwen, Focused QA alone reaches 65.6\%, but adding Overall Feedback improves alignment with expert reviews and human preference.
Both models improve expert-review alignment over pretrained baselines and are significantly preferred for usefulness (92.0\% and 67.3\%, respectively).
Demo examples are available online.\footnote{\url{https://hayeonbang.github.io/PitchIn}}

% [original]
% Effective vocal coaching must connect what needs improvement in a singing performance with how to improve it.
% Training such systems is challenging because recordings paired with expert diagnoses and corrective advice are scarce.
% Online singing communities provide a potential source of supervision, but observations and advice are scattered across unstructured comments.
% We introduce \textbf{PitchIn}, an audio--language model for natural-language vocal coaching, and construct the \textbf{PeerCoach Dataset} with 14,445 instruction examples covering Focused QA, Overall Feedback, and Recognition QA.
% We compare combinations of these target types on Qwen2.5-Omni and fine-tune both Qwen2.5-Omni and MOSS-Music on Focused QA and Overall Feedback.
% On VocalCoachBench, the proportion of requested diagnosis--correction pairs that both match an expert issue and give valid advice increases from 21.3\% to 61.0\% for Qwen2.5-Omni and from 31.4\% to 51.2\% for MOSS-Music.
% Training Qwen2.5-Omni directly on the source comments yields 19.8\%, no better than the pretrained model.
% Both PitchIn models also improve alignment with expert reviews.
% Human evaluators significantly prefer PitchIn responses to pretrained ones for usefulness at rates of 92.0\% and 67.3\%.
% Demo examples are available online.\footnote{\url{https://hayeonbang.github.io/PitchIn}}

% [revised]
\end{abstract}

\section{Introduction}
\label{sec:intro}

% Recent advances in audio--language models have improved their ability to understand audio and respond to questions in natural language~\cite{salmonn,qwen25omni}.
% However, helping users make judgments or develop skills in specialized domains requires more than understanding and describing the audio.
% In generating free-form coaching feedback, models must select task-relevant information and organize it into explanations and actionable advice suited to the user's goals~\cite{jiang2023expert}.
% Learning this capability requires audio annotated with specific diagnoses and advice.
% Yet detailed performance descriptions are difficult to obtain and annotate consistently~\cite{llaqo}.
% % Reconstructing existing feedback into task-appropriate training responses therefore offers a route to adapt audio--language models to specialized domains.

Recent advances in audio--language models have improved their ability to understand audio and respond to questions in natural language~\cite{salmonn,qwen25omni}.
However, supporting evaluation and skill development in specialized domains requires more than simply understanding or describing the audio.
Music education is one such setting, where a model must identify performance issues, explain them in context, and provide specific guidance for improvement~\cite{jiang2023expert}.
Learning to generate such feedback requires audio paired with detailed diagnoses and corrective advice, yet such annotations are difficult to obtain consistently~\cite{lerch2019music,llaqo}.

We study this problem in the setting of natural-language vocal coaching. Given a singing recording and a request for overall feedback, a model should identify salient strengths and weaknesses and provide actionable suggestions for improvement without being told which attributes to assess. A recent benchmark, VocalCoachBench (VCB), evaluates this capability against expert diagnoses and corrective guidance~\cite{vocalcoachbench}. We construct training supervision from online community feedback and adapt audio--language models to this task.

%We investigate this problem in the natural-language vocal coaching scenario. Given a singing recording and a request for overall feedback, a model should identify salient strengths and weaknesses, and provide actionable suggestions for improvements without being told which attributes to assess. A recent benchmark, VocalCoachBench (VCB) evaluates this capability against expert diagnoses and corrective guidance~\cite{vocalcoachbench}. The present study constructs training data for this task and adapts a model to it.

% [original]
% We draw on authentic interactions in online singing communities that link recordings, feedback requests, and corresponding observations and advice.
% Useful content is distributed across comments and varies in specificity and supporting evidence.
% We therefore select source-supported information from this corpus and use it to construct two forms of coaching supervision.
% Focused QA (FQA) extracts specific observations and advice into individual question--answer pairs, while Overall Feedback (OF) integrates distributed feedback into a complete response to an overall request.
% Recognition QA (RQ) supplements these coaching targets with multiple-choice questions supported by repeated community feedback or audio transformations.
% These three target types form the \textbf{PeerCoach Dataset}, 14,445 instruction examples built from 7,042 source recordings and covering 133.1 hours of input audio.

% [revised]

We draw on online singing-community interactions linking singing recordings, feedback requests in the original posts, and observations and advice in the source comments.
We turn these observations and advice into two types of coaching targets: Focused QA (FQA), which converts specific observations and advice into individual question--answer pairs, and Overall Feedback (OF), which integrates distributed feedback into a coherent response to an overall feedback request. 
Recognition QA (RQ) supplements these coaching targets with multiple-choice questions based on issues supported by multiple source comments or audio transformations. 
Together, these three target types form the \textbf{PeerCoach Dataset}, comprising 14,445 instruction examples from 7,042 source recordings and 133.1 hours of audio.

%We draw on authentic interactions in online singing communities that link recordings, feedback requests, and corresponding observations and advice. Useful content is distributed across comments and varies in specificity and supporting evidence. We therefore select source-supported information from this corpus and use it to construct two forms of coaching supervision. Focused QA (FQA) extracts specific observations and advice into individual question--answer pairs, while Overall Feedback (OF) integrates distributed feedback into a complete response to an overall request. Recognition QA (RQ) supplements these coaching targets with multiple-choice questions based on issues explicitly mentioned in multiple source comments or on audio transformations. These three target types form the \textbf{PeerCoach Dataset}, 14,445 instruction examples built from 7,042 source recordings and covering 133.1 hours of input audio.
% Each example combines a singing audio, an instruction, and a target response.

% [original]
% We fine-tune Qwen2.5-Omni \cite{qwen25omni} on different combinations of these target types to study their effects on free-form coaching and structured singing assessment. The FQA+OF models based on Qwen2.5-Omni and MOSS-Music \cite{mossmusic} are denoted PitchIn-Q and PitchIn-M, respectively, and collectively as \textbf{PitchIn}. Figure~\ref{fig:vocalcoach_overview} summarizes the dataset construction and adaptation pipeline.

\begin{figure*}[t]
  \centering
  \includegraphics[width=0.98\textwidth]{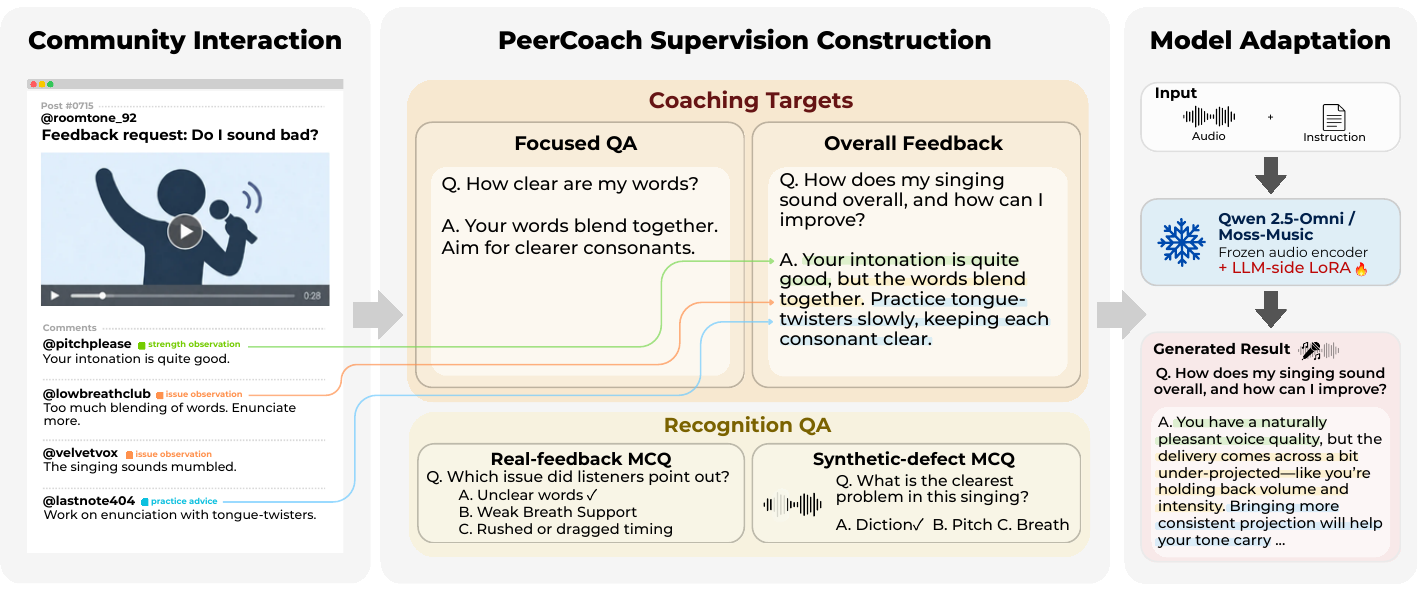}
  \vspace{-3pt}
  \caption{Overview of PitchIn.
  Community singing recordings and feedback are used to construct Focused QA, Overall Feedback, and auxiliary Recognition QA targets for audio--language model adaptation.}
  \label{fig:vocalcoach_overview}
\end{figure*}

% [revised]
We fine-tune Qwen2.5-Omni \cite{qwen25omni} on different combinations of these target types to study their effects on free-form coaching and structured singing assessment. 
The FQA+OF models based on Qwen2.5-Omni and MOSS-Music are denoted PitchIn-Q and PitchIn-M, respectively, and collectively \textbf{PitchIn}.
Figure~\ref{fig:vocalcoach_overview} summarizes the dataset construction and adaptation pipeline.

% [original]
% Our work makes three contributions. First, we introduce the PeerCoach Dataset and a method for constructing coaching targets from observations and advice distributed across community comments. Second, we develop PitchIn and show that FQA+OF fine-tuning improves alignment with expert reviews and the success rate of diagnosis--correction pairs over the pretrained baselines of both backbones. Finally, we show that structured OF supervision outperforms direct training on source comments and analyze the complementary effects of OF and RQ.

% [revised]
Our work makes three contributions. First, we introduce the PeerCoach Dataset and a method for constructing coaching targets from observations and advice distributed across community comments. Second, we develop PitchIn and show that FQA+OF fine-tuning improves both alignment with expert reviews and the proportion of requested diagnosis--correction pairs that match an expert issue with valid advice over the pretrained baselines of both backbones. Finally, we show that structured OF supervision outperforms direct training on source comments and analyze the effects of adding OF and RQ.

\vspace{-10pt}
\section{Related Work}
\label{sec:related}
Automatic singing assessment has largely focused on predicting quality scores or rankings from attributes such as pitch, rhythm, and timbre~\cite{gupta2020leaderboard}. VocalVerse extends this setting to descriptive feedback~\cite{vocalverse}, but its evaluation remains centered on the perceptual validity of quality rankings. In contrast, we target coaching-oriented diagnosis and corrective advice and evaluate them against expert reviews using VocalCoachBench~\cite{vocalcoachbench}.

Prior work constructs music instruction data from captions, tags, and annotations: MU-LLaMA builds MusicQA from captions and tags~\cite{mullama}, LLark augments annotations with extracted tempo and chords~\cite{llark}, and Music Flamingo combines captions and metadata into detailed descriptions and skill-focused QA~\cite{musicflamingo}. In contrast, we construct coaching targets from distributed, unstructured community feedback tied to individual performances.

% Prior work on constructing music instruction datasets converts existing captions, tags, and annotations into training examples for audio--language models. MU-LLaMA generates MusicQA's open-ended question--answer pairs from music captions and tags~\cite{mullama}. LLark augments annotations with automatically extracted information, including tempo and chords~\cite{llark}. Music Flamingo combines initial captions and metadata into detailed music descriptions and skill-focused QA~\cite{musicflamingo}.

% [original]
% For music coaching, LLaQo converts teachers' free-response feedback into question--answer pairs using an assessment rubric~\cite{llaqo}.
% It combines these pairs with performance annotations and ratings to train a query-based coach.
% This provides a precedent for constructing coaching supervision from written performance feedback.
% Its source is expert feedback written for one performance, whereas community feedback on a recording is spread across the comments of many listeners.
% We examine whether constructing these comments into coaching targets improves generated feedback compared with training on the original comments.

% [revised]
For music coaching, LLaQo converts teachers' free-response feedback into question--answer pairs using an assessment rubric~\cite{llaqo}.
It combines these pairs with performance annotations and ratings to train a query-based coach.
LLaQo uses expert feedback on individual performances, whereas community feedback on each recording spans multiple listeners' comments.
We study whether converting these comments into vocal coaching targets improves coaching responses over training directly on the source comments.

% We examine whether targets built this way train better coaching than the original comments.

\section{The PeerCoach Dataset}
\label{sec:data}

% To turn community discussions into training supervision, we organize their observations and advice into specific coaching answers and integrated overall responses.
% We also construct auxiliary recognition questions from repeated feedback and controlled audio transformations, allowing us to examine how these different targets contribute to coaching and assessment.

\subsection{Community Corpus}
% [original]
% We collected singing-feedback posts and their comment threads from Reddit's r/singing\footnote{\url{https://www.reddit.com/r/singing/}} and r/ratemysinging\footnote{\url{https://www.reddit.com/r/ratemysinging/}} communities over a 40-month period.
% We trimmed each recording tosegments containing singing, excluded those longer than 180 s after trimming, and extracted vocal stems using Mel-Band RoFormer~\cite{melbandroformer}. 
% The corpus contains 9,753 singing recordings, 75,684 comments, and 158.1 hours of audio.

% [revised]
We collected singing-feedback posts and their comment threads from Reddit's r/singing\footnote{\url{https://www.reddit.com/r/singing/}} and r/ratemysinging\footnote{\url{https://www.reddit.com/r/ratemysinging/}} communities over a 40-month period.
We trimmed each recording to segments containing singing, excluded recordings longer than 180 s after trimming, and extracted vocal stems from the retained trimmed recordings using Mel-Band RoFormer~\cite{melbandroformer}.
The corpus contains 9,753 singing recordings, 75,684 comments, and 158.1 hours of audio.

\vspace{-5pt}
\subsection{Coaching Targets}

We construct two kinds of coaching target from the original posts and observations and advice from the source comments. 
Most posts offer only a few specific observations, and far fewer offer enough for a complete assessment.
Each training example pairs a singing recording and a question with an English response addressed directly to the singer; neither question nor response may introduce unsupported diagnoses, causes, or confident claims.
%Neither prompt may state a diagnosis, a cause, or a confident claim that the comments do not support.

\noindent\textbf{Focused QA (FQA).}
FQA turns individual observations into questions and answers.
GPT-5.2~\cite{openai_models} generates up to three such pairs per post.
Questions preserve the content of the original post and may ask for general feedback or target a specific aspect of singing.
Each answer is grounded only in comments from users other than the poster.
% Each answer must be grounded in comments written by users other than the poster.
% When a question asks for general feedback, its answer covers at most two observations.
Technical diagnoses require concrete observational cues, and examples are discarded when the source comments provide insufficient support or are irrelevant to the question. 
Brief practice advice is added only when supported by a clearly identified issue. 

% [original]
% \noindent\textbf{Overall Feedback (OF).}
% OF composes a coherent response from observations and advice distributed across a thread. 
% A feedback item is a single observation or piece of practice advice extracted from the source comments.
% We use GPT-OSS-120B~\cite{gptoss120b} to categorize feedback items as strength observations, issue observations, or practice advice.
% Eligible posts require at least three feedback items, including an issue observation and either a strength observation or practice advice.
% We exclude posts with only generic feedback or feedback inapplicable to isolated vocals.
% GPT-5.5 answers the fixed question \emph{How does my singing sound overall, and how can I improve?}, merging repeated observations and advice while preserving distinct observations and concrete practice advice without fixing their number.
% When no practice advice is available, it may add a brief action directly implied by the observations.
% Rule-based filters remove 415 responses with potentially unsupported detail, stock phrasing, or references to the comments, leaving 1,909 responses before held-out exclusion.

% [revised]
\noindent\textbf{Overall Feedback (OF).}
OF composes a coherent response from observations and advice distributed across a thread. 
A feedback item is a single observation or piece of practice advice extracted from the source comments.
We use GPT-OSS-120B~\cite{gptoss120b} to categorize feedback items as strength observations, issue observations, or practice advice.
Eligible posts require at least three feedback items, including an issue observation and either a strength observation or practice advice.
We exclude posts with only generic feedback or feedback inapplicable to isolated vocals.
GPT-5.5 answers the fixed question \emph{How does my singing sound overall, and how can I improve?}, merging repeated observations and advice while preserving distinct observations and concrete practice advice without fixing their number.
When no practice advice is available, it may add a brief action directly implied by the observations.
Rule-based filtering leaves 1,909 responses after removing 415 with potentially unsupported detail, stock phrasing, or comment references; excluding responses from 190 reserved posts yields 1,719 OF examples.

\noindent\textbf{Source Fidelity Audit.}
GPT-5.6-sol audits 100 FQA/OF target responses against their source comments, using a stratified sample of 50 posts from an evidence-rich subset of 723 posts.
Among claims requiring support from the source comments, 93.6\% are fully supported for FQA and 85.5\% for OF.
Actionable advice appears in 80\% of FQA responses and 98\% of OF responses.

\noindent\textbf{Listening Audit.} One author with music expertise audited 100 responses from 50 posts randomly sampled from the 1,672 with both FQA and OF targets.
At the response level, 72\% of FQA and 84\% of OF targets were judged fully supported by recordings; 16\% and 10\%, respectively, were partially supported.
Denominators include one indeterminate response per target type.

\vspace{-3pt}
\subsection{Auxiliary Recognition Targets}
\vspace{-3pt}
% [original]
% Alongside the free-form coaching targets, we construct multiple-choice \textbf{Recognition QA (RQ)} that requires the model to identify a specific singing problem. 

% [revised]
We construct multiple-choice \textbf{Recognition QA (RQ)} as auxiliary supervision to test whether explicit singing-problem recognition improves open-ended diagnosis.

\noindent\textbf{Real-Feedback MCQ.} We select issue observations explicitly mentioned in at least two distinct comments. 
GPT-5.5 verifies that the source comments refer to the same problem; the supported problem becomes the correct choice, while categories not mentioned in the source comments serve as distractors. 
Each target response includes the selected option and a brief explanation.

% [original]
\noindent\textbf{Synthetic-Defect MCQ.}
We synthesize four defects: pitch inaccuracy by segment-wise detuning (up to 80 cents), pitch instability by 0.8--1.5 Hz modulation (35--65 cents), timing irregularity by local time stretching (up to 12\%), and dynamic imbalance by segment gains of 0.3--2.4. Each transformation affects 1--5 spans of 2--8 s, and its type defines the target label. To prevent processing artifacts from becoming label cues, we include null controls using micro-transformations or a transformation followed by its inverse. Table~\ref{tab:coaching_data} summarizes the dataset.

% [revised]
% \noindent\textbf{Synthetic-Defect MCQ.}
% We synthesize pitch inaccuracy (segment-wise detuning up to 80 cents), pitch instability (0.8--1.5 Hz modulation, 35--65 cents), timing irregularity (local time stretching up to 12\%), and dynamic imbalance (segment gains of 0.3--2.4). Transformations affect 1--5 spans of 2--8 s; their types define target labels. Null controls use micro-transformations or transformations followed by their inverses to prevent artifact-based label cues. Table~\ref{tab:coaching_data} summarizes the dataset.

%Segment-wise detuning of up to 80 cents produces pitch inaccuracy, and a 0.8 to 1.5 Hz modulation of 35 to 65 cents produces pitch instability. Local time stretching by up to 12 percent produces timing irregularity, and segment gains between 0.3 and 2.4 produce dynamic imbalance. Each segment-wise transformation covers one to five spans of two to eight seconds. The applied transformation provides the target label. To keep processing artifacts from acting as label cues, we include null controls with micro-transformations or a transformation followed by its inverse. Table~\ref{tab:coaching_data} summarizes the dataset.

\begin{table}[t]
\centering
\caption{PeerCoach Dataset size by target type. Sources count unique recordings; hours sum durations of unique input files, including transformed audio. Totals are deduplicated across types.}
\label{tab:coaching_data}
\scalebox{0.98}{%
\begin{tabular}{@{}lrrr@{}}
\toprule
Type & Sources & Examples & Hours \\
\midrule
Focused QA (FQA) & 6,979 & 10,919 & 114.2 \\
Overall Feedback (OF) & 1,719 & 1,719 & 28.5 \\
Real-Feedback MCQ (RQ) & 393 & 411 & 6.2 \\
Synthetic-Defect MCQ (RQ) & 466 & 1,396 & 18.2 \\
\midrule
Total & 7,042 & 14,445 & 133.1 \\
\bottomrule
\end{tabular}
}
\end{table}

\begin{table*}[!t]
\centering
\caption{Results for PitchIn and comparison models (scores $\times100$).
Higher is better except for Self-B4.
Shaded rows identify the two PitchIn models.
Bold scores mark the best displayed value within the Qwen and MOSS groups, including ties.
$\dagger$ indicates a significant difference from the corresponding pretrained baseline after joint Holm correction over 88 tests ($p<0.05$).}
\label{tab:supervision_results}
\begingroup
\small
\setlength{\tabcolsep}{5.5pt}
\renewcommand{\arraystretch}{1.12}
\scalebox{0.95}{%
\begin{tabular}{@{}lrrrrrrrrrrrr@{}}
\toprule
& \multicolumn{4}{c}{\textbf{Natural Overall}} & \multicolumn{4}{c}{\textbf{VCB Open-ended}} & \multicolumn{4}{c}{\textbf{VCB Structured}} \\
\cmidrule(lr){2-5}\cmidrule(lr){6-9}\cmidrule(lr){10-13}
& \multicolumn{3}{c}{Reference Alignment} & Repetition & \multicolumn{2}{c}{Diagnosis} & Correction & \multirow{2}{*}{E2E} & \multicolumn{2}{c}{Ranking} & Top-3 & Segment \\
\cmidrule(lr){2-4}\cmidrule(lr){5-5}\cmidrule(lr){6-7}\cmidrule(lr){8-8}\cmidrule(lr){10-11}\cmidrule(lr){12-12}\cmidrule(lr){13-13}
\textbf{Model} & B-U & SS & SPICE & Self-B4$\downarrow$ & S.Hit & Hit & Valid & & Direct & Score & m-F1@3 & Match \\
\specialrule{\heavyrulewidth}{\aboverulesep}{\belowrulesep}
\multicolumn{13}{@{}l@{}}{\textit{Qwen2.5-Omni}} \\
Pretrained & 2.2 & 51.2 & 6.3 & 85.4 & 3.1 & \textbf{63.4} & 31.8 & 21.3 & 68.7 & 4.4 & 31.1 & 30.9 \\
Raw Comments & 3.0 & 54.3 & 5.4$^{\dagger}$ & \textbf{72.2} & 2.1 & 42.2$^{\dagger}$ & 31.3 & 19.8 & \textbf{71.0} & 58.4$^{\dagger}$ & 30.0 & 32.4 \\
\rowcolor[gray]{0.95}
\textbf{PitchIn-Q} & \textbf{14.6}$^{\dagger}$ & \textbf{71.2}$^{\dagger}$ & \textbf{10.3}$^{\dagger}$ & 77.3 & \textbf{6.8}$^{\dagger}$ & 48.1$^{\dagger}$ & 98.2$^{\dagger}$ & 61.0$^{\dagger}$ & 64.6 & \textbf{62.0}$^{\dagger}$ & \textbf{32.3} & \textbf{34.0} \\
\addlinespace[2pt]
\hspace{5pt}FQA only & 7.7$^{\dagger}$ & 64.4$^{\dagger}$ & 7.5$^{\dagger}$ & 81.5 & 6.4$^{\dagger}$ & 52.5$^{\dagger}$ & \textbf{98.9}$^{\dagger}$ & \textbf{65.6}$^{\dagger}$ & 64.5 & 51.1$^{\dagger}$ & 27.9$^{\dagger}$ & 32.4 \\
\hspace{5pt}FQA+OF+RQ & 13.2$^{\dagger}$ & 69.6$^{\dagger}$ & 9.9$^{\dagger}$ & 76.5 & \textbf{6.8}$^{\dagger}$ & 50.5$^{\dagger}$ & 93.3$^{\dagger}$ & 58.9$^{\dagger}$ & 69.2 & 57.0$^{\dagger}$ & 30.4 & 32.1 \\
\midrule
\multicolumn{13}{@{}l@{}}{\textit{MOSS-Music}} \\
Pretrained & 10.8 & 46.8 & 10.4 & 74.0 & 2.7 & 25.5 & 52.0 & 31.4 & 61.5 & 15.7 & \textbf{29.0} & 32.1 \\
\rowcolor[gray]{0.95}
\textbf{PitchIn-M} & \textbf{14.0}$^{\dagger}$ & \textbf{60.4}$^{\dagger}$ & \textbf{11.1}$^{\dagger}$ & \textbf{70.3} & \textbf{5.2}$^{\dagger}$ & \textbf{44.7}$^{\dagger}$ & \textbf{86.6}$^{\dagger}$ & \textbf{51.2}$^{\dagger}$ & \textbf{70.5}$^{\dagger}$ & \textbf{45.9}$^{\dagger}$ & 25.7$^{\dagger}$ & \textbf{32.8} \\
\midrule
\multicolumn{13}{@{}l@{}}{\textit{External comparison models}} \\
Fun-Audio-Chat & 9.6 & 60.3 & 9.4 & 80.5 & 5.4 & 57.7 & 66.7 & 47.1 & 48.8 & 31.6 & 29.7 & 33.6 \\
MiMo-Audio & 5.1 & 53.7 & 7.8 & 91.1 & 3.0 & 39.2 & 84.7 & 42.7 & 64.5 & 36.9 & 30.1 & 40.5 \\
\bottomrule
\end{tabular}
}
\endgroup
\par\smallskip
\begin{minipage}{\linewidth}
\footnotesize
Pretrained Qwen returns all 2,575 requested pairs, whereas pretrained MOSS leaves 1,190 empty.
Metric definitions are in Sec.~\ref{sec:evaluation_protocol}.
\end{minipage}
\end{table*}

\section{Experimental Design}
\subsection{Models and Adaptation}
\label{sec:design}

We fine-tune Qwen2.5-Omni-7B~\cite{qwen25omni} on FQA+OF as \textbf{PitchIn-Q}, comparing it with pretrained, FQA-only, and FQA+OF+RQ variants to assess OF and RQ.
FQA+OF contains 11,400 training and 1,238 validation examples, split by input audio file using a fixed seed, with assignments shared across configurations.
RQ transformations cause 165 source recordings to span both splits in FQA+OF+RQ.
To assess target construction, the \textit{Raw Comments} control uses the source comments underlying FQA and OF, matching PitchIn-Q's recordings, example exposures, and optimizer updates.
All fine-tuned Qwen variants in Table~\ref{tab:supervision_results} start independently from the same pretrained checkpoint.
We also fine-tune MOSS-Music-8B-Instruct~\cite{mossmusic} on FQA+OF as \textbf{PitchIn-M}, testing these targets on a model for music understanding with its native processor and chat format.
It shares PitchIn-Q's examples and split; pretrained MOSS-Music is its baseline.
External baselines are comparably sized open-weight audio--language models Fun-Audio-Chat-8B~\cite{funaudiochat} and MiMo-Audio-7B-Instruct~\cite{mimoaudio}.
% We focus on FQA+OF for free-form coaching because it achieves higher Natural Overall alignment than FQA-only, although FQA-only achieves higher E2E on VCB Open-ended.
%We compare PitchIn-Q with its pretrained backbone and with FQA and FQA+OF+RQ variants to examine the effects of OF and RQ. 
%We preset its training duration to match PitchIn-Q's example exposures and optimizer updates.
%All fine-tuned Qwen models are independently initialized from the same pretrained checkpoint.
% We select Fun-Audio-Chat-8B~\cite{funaudiochat} and MiMo-Audio-7B-Instruct~\cite{mimoaudio} as open-weight references with parameter counts comparable to PitchIn.

We mix target types without reweighting or oversampling and train rank-64 LoRA adapters~\cite{lora} on the language model's attention and MLP projections while freezing all pretrained parameters, including the audio encoder.
%We mix examples from the target types, without type-level reweighting or oversampling. We train rank-64 LoRA adapters~\cite{lora} in the language model's attention and MLP projections while freezing all pretrained parameters, including the audio encoder.
% The objective is response-token negative log-likelihood with prompts and padding masked.
AdamW~\cite{adamw} uses a peak learning rate of $2\times10^{-5}$, an effective batch size of 16, and a cosine schedule with 1,250 warm-up updates over a 25,000-update horizon.
%AdamW~\cite{adamw} uses a peak learning rate of $2\times10^{-5}$ and an effective batch size of 16. The cosine schedule has a 25,000-update horizon and 1,250 warm-up updates.
Audio is resampled to 16 kHz and zero-padded to 180 s. For PeerCoach-trained models, we select the completed-epoch checkpoint with the lowest validation loss; epoch 4 is selected for all configurations.
%Audio is resampled to 16 kHz and zero-padded to 180 seconds. For models trained on PeerCoach targets, we select the checkpoint with the lowest validation loss among completed epochs. This selects epoch 4 for each configuration.

\subsection{Evaluation Protocol}
\label{sec:evaluation_protocol}

% [original]
% We evaluate on VCB, which pairs 515 singing recordings with 1,064 expert reviews~\cite{vocalcoachbench}.
% The recordings come from ten public singing datasets.
% VCB recordings and annotations are excluded from fine-tuning.
% We evaluate three separately prompted tasks.
% Natural Overall scores the complete response, while the two VCB tasks use their own prompts and fixed output formats to make individual claims scorable.
% Both VCB tasks use greedy decoding, and missing or unusable predictions count as failures.

% [revised]
We evaluate on VCB, which pairs 515 singing recordings from ten public datasets with 1,064 expert reviews~\cite{vocalcoachbench}.
VCB recordings and annotations are excluded from fine-tuning.
Three tasks use separate prompts: Natural Overall scores complete responses, while the two VCB tasks use fixed output formats for scoring model outputs.
Both VCB tasks use greedy decoding; missing or unusable predictions count as failures.

% [original]
% Models answer \emph{How does my singing sound overall, and how can I improve?} with a single response.
% Each recording carries about two expert reviews, and alignment is scored against all of them.
% B-U (mean corpus BLEU-1--4)~\cite{bleu} measures wording overlap, sentiment similarity (SS) measures agreement in stance, and SPICE~\cite{spice} measures semantic content.
% Self-BLEU-4~\cite{selfbleu} measures how much a model repeats itself across recordings, so we report it as a diagnostic and exclude it from the paired tests.
% All models receive full recordings and the same question in their native system formats.
% Decoding uses temperature 0.6, top-p 0.85, top-k 30 and repetition penalty 1.08.
% Responses are limited to 384 new tokens under recording-specific fixed seeds.

% [revised]
\noindent\textbf{Natural Overall.} Models answer \emph{How does my singing sound overall, and how can I improve?} with a single response.
Alignment is scored against all expert reviews (about two per recording).
B-U (mean corpus BLEU-1--4)~\cite{bleu} measures wording overlap, sentiment similarity (SS) measures agreement in stance, and SPICE~\cite{spice} measures semantic content.
Self-BLEU-4~\cite{selfbleu} measures repetition across recordings and is reported as a diagnostic, excluded from paired tests.
All models receive full recordings in their native system formats.
Decoding uses temperature 0.6, top-p 0.85, top-k 30 and repetition penalty 1.08.
Responses are limited to 384 new tokens under recording-specific fixed seeds.

% [original]
% Following VCB~\cite{vocalcoachbench}, models generate five diagnosis--correction pairs per recording in a fixed schema.
% We evaluate these pairs using GPT-OSS-120B~\cite{gptoss120b}.
% Of the 6,162 expert issue-claims, S.Hit counts those matched by the same specific problem and Hit also counts those matched only by the same general issue.
% Valid measures whether correction is specific, actionable, and appropriate for its linked diagnosis.
% The primary metric, E2E, requires a diagnosis matching an expert issue at either level and valid linked advice.
% Valid and E2E use all five requested slots per recording.

% [revised]
\noindent\textbf{VCB Open-ended.} Following VCB~\cite{vocalcoachbench}, models generate five diagnosis--correction pairs per recording in a fixed schema, evaluated by GPT-OSS-120B~\cite{gptoss120b}.
Among 6,162 expert diagnoses, S.Hit counts matches at the specific problem level; Hit also accepts general issue matches.
Valid measures the fraction of requested pairs with specific, actionable corrections appropriate to their diagnoses.
E2E additionally requires a diagnosis matching an expert issue at either level.
Both use all 2,575 requested pairs, counting unfilled pairs as failures.
Coverage measures the fraction of expert diagnoses matched with valid linked corrections, counting each once and reporting matches at either level and strict matches separately.

% [original]
% We follow VCB~\cite{vocalcoachbench} to evaluate issue-category prediction and performance ranking.
% Top-3 measures macro-F1 across seven issue categories using the three categories predicted for each recording.
% Segment Match measures the proportion of 262 segments whose predicted issue category matches any expert label.
% Direct Ranking asks models to compare two recordings.
% Score-derived Ranking orders independently predicted 0--5 quality scores.
% Both ranking metrics measure pairwise agreement over 567 pairs from 189 triplets, each scored against both expert reference orders.
% Tied predicted scores receive no credit.

% [revised]
\noindent\textbf{VCB Structured.} We follow VCB~\cite{vocalcoachbench} to evaluate issue-category prediction and performance ranking.
Top-3 measures macro-F1 over seven issue categories from three predictions per recording.
Segment Match is the proportion of 262 segments with category predictions matching any expert label.
Direct Ranking compares two recordings; Score-derived Ranking orders independently predicted 0--5 quality scores.
Both measure pairwise agreement on 567 pairs from 189 triplets against both expert reference orders.
Tied predicted scores receive no credit.

% Synthetic recognition protocol is described with the results in Section 5.2.
% \noindent\textbf{Synthetic Recognition.}
% We apply the four RQ transformations to each of 300 recordings excluded from fine-tuning and validation, with settings fixed on a separate discovery set.
% For each pair of transformed versions, a match counts when the model's option logits favor the correct label assignment over the swapped one, so chance is 50\%.
% All versions of a recording are trimmed to a common length so that duration cannot reveal time stretching.

\noindent\textbf{Statistical analysis.}
Two-sided paired randomization tests~\cite{yeh2000} preserve recording, segment-source, or ranking-triplet units.
We jointly apply Holm correction~\cite{holm1979} to 88 tests covering eight within-backbone contrasts across 11 metrics.
These contrasts compare five fine-tuned models with their respective pretrained backbones and PitchIn-Q with Raw Comments, FQA-only, and FQA+OF+RQ.

\vspace{-5pt}
\subsubsection{Human Evaluation}
\vspace{-5pt}
Ten evaluators with formal music training rate pairs of model-blinded Natural Overall responses on a five-point preference scale for usefulness and recording fit.
Contrasts compare PitchIn-Q with pretrained Qwen and FQA-only, and PitchIn-M with pretrained MOSS.
Each evaluator judges seven pairs per contrast with Korean translations and counterbalanced positions (210 judgments, 22 recordings).
Preference rates pool slight and strong preferences, weight recordings equally and keep ties.
Exact sign-flip tests assess net preference by recording and evaluator, with separate Holm corrections over six tests each.

\section{Results and Analysis}
\label{sec:results}

\subsection{Coaching Gains on Two Backbones}
\vspace{-3pt}
% [original]
% PitchIn-Q and PitchIn-M align more closely with expert reviews than their pretrained baselines, with significant gains on all three Natural Overall reference-alignment metrics (Table~\ref{tab:supervision_results}).
% E2E improves significantly from 21.3\% to 61.0\% for PitchIn-Q and from 31.4\% to 51.2\% for PitchIn-M.
% PitchIn-Q's E2E gain accompanies an increase in correction validity from 31.8\% to 98.2\%.
% Its diagnoses grow from average 4.7 to 13.0 words and cover fewer expert issues, with Hit falling from 63.4\% to 48.1\% while S.Hit rises from 3.1\% to 6.8\%.
% On one recording, pretrained Qwen's \textit{``The pronunciation is unclear''} matches four expert claims, including those on consonant--vowel timing and tongue position.
% PitchIn-Q's \textit{``Words are hard to understand, especially at the start of phrases, likely due to breathiness''} matches only the two on broken diction.
% Counting each expert issue-claim once, the share matched by a diagnosis with a valid correction rises from 19.0\% to 47.9\% for PitchIn-Q, and from 1.5\% to 6.8\% for strict matches.
% Both PitchIn models also outperform Fun-Audio-Chat and MiMo-Audio on reference alignment, and PitchIn-Q does so on E2E as well.

% [revised]
Both PitchIn models significantly outperform their pretrained baselines on all three Natural Overall alignment metrics (Table~\ref{tab:supervision_results}).
E2E improves significantly from 21.3\% to 61.0\% for PitchIn-Q and from 31.4\% to 51.2\% for PitchIn-M.
PitchIn-Q's correction validity rises from 31.8\% to 98.2\%, and its diagnoses lengthen from 4.7 to 13.0 words on average, while Hit falls from 63.4\% to 48.1\% and S.Hit rises from 3.1\% to 6.8\%.
On one recording, pretrained Qwen's \textit{``The pronunciation is unclear''} matches four expert diagnoses, including those on consonant--vowel timing and tongue position.
PitchIn-Q's \textit{``Words are hard to understand, especially at the start of phrases, likely due to breathiness''} matches only the two on broken diction.
Counting each expert diagnosis once, PitchIn-Q's coverage with valid linked corrections rises from 19.0\% to 47.9\% (strict matches: 1.5\% to 6.8\%).
Both PitchIn models also outperform Fun-Audio-Chat and MiMo-Audio on Natural Overall alignment and E2E.

% [original]
% Although the coaching targets contain virtually no numeric ratings, Score-derived Ranking rises from 4.4\% to 62.0\% for Qwen and from 15.7\% to 45.9\% for MOSS.
% Pretrained Qwen assigns 2.5 to 502 of 515 recordings, leaving 525 of the 567 pairs tied.
% PitchIn-Q resolves 398 of these ties.
% Its score order on these pairs agrees with 80.7\% of expert judgments, against 72.1\% for pretrained Qwen's direct comparisons.
% Top-3 and segment-level predictions remain concentrated on a few categories for all models.

% [revised]
Despite virtually no numeric ratings in the coaching targets, Score-derived Ranking rises from 4.4\% to 62.0\% for Qwen and from 15.7\% to 45.9\% for MOSS.
Pretrained Qwen assigns 2.5 to 502 of 515 recordings, tying 525 of 567 pairs.
Raw Comments also reaches 58.4\%, showing that much of the improvement over pretrained Qwen is not specific to constructed coaching targets.
PitchIn-Q resolves 398 ties, agreeing with expert reference orders in 80.7\% of judgments on these pairs, versus 72.1\% for pretrained Qwen's direct comparisons.
All models' Top-3 and Segment Match predictions remain concentrated on a few categories.

\subsection{Contributions of the PeerCoach Target Types}
PitchIn-Q significantly outperforms Raw Comments on all Natural Overall alignment and VCB Open-ended metrics (E2E: 61.0\% versus 19.8\%).
From the same FQA checkpoint, OF outperforms source comments on all three alignment metrics under matched recordings, questions, example exposures, and optimizer updates.

% [original]
% Adding OF brings coaching responses closer to expert reviews and improves score-derived ranking.
% All Natural Overall alignment metrics improve significantly over FQA, and FQA repeated for the same example exposures and optimizer updates still trails on all three.
% Score-derived Ranking rises significantly from 51.1\% to 62.0\%, with tied pairs decreasing from 206 to 134.
% On the VCB Open-ended task, FQA-only achieves significantly higher E2E than FQA+OF (65.6\% versus 61.0\%).
% The lower E2E with OF accompanies lower Hit, while correction validity remains above 98\% for both models.

% [revised]
OF significantly improves all three alignment metrics over FQA-only, which still trails on all three with matched example exposures and optimizer updates.
Score-derived Ranking rises significantly from 51.1\% to 62.0\%, with ties falling from 206 to 134.
FQA-only has significantly higher VCB Open-ended E2E than FQA+OF (65.6\% versus 61.0\%), with lower Hit for FQA+OF; correction validity exceeds 98\% for both.

% [original]
% Adding RQ yields no significant gains in Natural Overall alignment or VCB Open-ended performance.
% Correction validity decreases significantly, while Direct Ranking improves significantly.
% We therefore examine whether RQ improves recognition of the synthetic transformations used in training on held-out recordings.
% Length-matched transformation--label matching rises from 49.4\% for PitchIn-Q to 62.8\% for FQA+OF+RQ (50\% chance).
% % Likelihood-based four-way MCQ accuracy increases from 24.7\% to 26.9\% (25\% chance).
% This gain remains significant after multiple-comparison correction.
% % Both gains are significant after multiple-comparison correction.

% [revised]
Adding RQ gives no significant Natural Overall alignment or VCB Open-ended gains; B-U, correction validity, and Top-3 macro-F1 decline significantly.
Direct Ranking improves significantly to near the pretrained level.
We apply all four RQ transformations to 300 recordings excluded from fine-tuning and validation, fixing settings on a separate discovery set and trimming each recording's versions to equal length.
Scores average all six transformation pairs per recording: 1 if summed option logits favor the correct assignment, 0 if they favor the swapped assignment, and 0.5 for exact ties.
On this contrastive matching probe (50\% chance), FQA+OF+RQ scores 62.8\%, significantly above pretrained Qwen (49.9\%) and PitchIn-Q (49.4\%).
The latter two remain near chance without a significant difference; their sensitivity to naturally occurring singing problems remains unresolved.
Recording-level exact sign-flip tests use Bonferroni (four original tests) and Holm (three pretrained follow-up tests) corrections.

% Previous wording of the auxiliary synthetic-defect matching analysis.
% With settings fixed on a separate discovery set, we apply all four RQ transformations to 300 recordings excluded from fine-tuning and validation, trimming each recording's transformed versions to equal length.
% For each pair of transformed versions, a match occurs when option logits favor the correct over the swapped label assignment.
% Match rates rise from 49.4\% (PitchIn-Q) to 62.8\% (FQA+OF+RQ; 50\% chance), with the gain remaining significant after multiple-comparison correction.

\begin{figure}[t]
  \centering
  \includegraphics[width=0.9\linewidth]{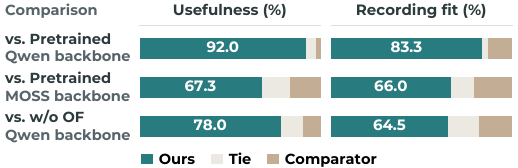}
\caption{Human preferences for usefulness and recording fit. Ours denotes PitchIn-Q and PitchIn-M.}
  \label{fig:human_preference}
\end{figure}

\vspace{-5pt}
\subsection{Human Evaluation}
Evaluators judge PitchIn's coaching responses as more useful and better fitted to the recording than those of each pretrained backbone (Fig.~\ref{fig:human_preference}).
PitchIn-Q's preference rates are 92.0\% for usefulness and 83.3\% for recording fit, and PitchIn-M's are 67.3\% and 66.0\%.
Evaluators prefer PitchIn-Q to FQA-only, at 78.0\% for usefulness and 64.5\% for recording fit, so the gain from OF extends to human judgments.
All six preferences are significant with recordings as the unit, and all but PitchIn-M's recording fit remain significant with evaluators as the unit.
Recording fit trails usefulness in all three contrasts.

% [original]
% With \textbf{PitchIn} and the \textbf{PeerCoach Dataset}, we show that community feedback data collected online can be used to significantly improve the performance of a vocal coaching system without manual expert annotation. In a situation where it is difficult to obtain large-scale training data for coaching supervision, this approach opens up a realistic possibility for implementing vocal coaching across diverse music domains.
% Despite these gains, strict matching to expert diagnoses remains limited, indicating substantial room for improvement in precise diagnosis.
% Future work will investigate multi-turn coaching and evaluate its effects on singers' learning across more diverse singing conditions.

% [revised]
\vspace{-7pt}
\section{Conclusion}
\label{sec:conclusion}
\vspace{-7pt}
We presented the \textbf{PeerCoach Dataset} and \textbf{PitchIn} for natural-language vocal coaching.
We showed that coaching targets constructed from community feedback improve vocal coaching without collecting expert coaching annotations for fine-tuning.
Strict matching to expert diagnoses nevertheless remains low. RQ improves performance on the synthetic recognition probe but yields no significant gains in open-ended diagnosis.
These results motivate testing whether timestamped issue annotations and priority labels improve audio-grounded diagnosis and issue prioritization.

\section{Data/Ethics statement}

% [original]
% We plan to release only reconstructed coaching text and links to the source audio, without redistributing original recordings or comments. Rights to the source content remain with their respective holders; we do not claim ownership of or relicense that content. We contacted Reddit regarding this research use but have not received a substantive response.

% [revised]
We plan to release only reconstructed coaching text and source audio links, without redistributing original recordings or posts.
Source content rights remain with their holders; we neither claim ownership nor relicense the content.
We officially contacted Reddit about this research use but have received no substantive response.

\noindent\textbf{LLM use.} We used large language models for data cleaning, constructing, improving language and clarity, accelerating code development and visualization, and related-work discovery and feedback on research ideas.

% But community text does not provide evidence about the particular recording, and converting it into recognition targets yields no detectable benefit from the recording itself. 
% Aligning coaching with the actual problems in each recording therefore remains an open challenge.

% We present PitchIn, an audio--language model for natural-language vocal coaching trained on supervision reconstructed from community feedback.
% Fine-tuning improves reference-text alignment and diagnosis--correction success over the respective pretrained baselines of Qwen2.5-Omni and MOSS-Music.
% Within Qwen, the preferred supervision composition depends on the coaching output: Focused QA gives the strongest diagnosis--correction success, whereas adding overall-feedback targets improves alignment of complete responses.
% Recognition targets yield more selective benefits, particularly for direct quality comparisons.
% Despite these gains, strict matching to expert diagnoses remains limited, indicating substantial room for improvement in precise diagnosis.
% Future work will investigate multi-turn coaching and evaluate its effects on singers' learning across more diverse singing conditions.

\bibliographystyle{IEEEbib}
\bibliography{refs}

@inproceedings{salmonn,
  author = {Tang, Changli and Yu, Wenyi and Sun, Guangzhi and Chen, Xianzhao and Tan, Tian and Li, Wei and Lu, Lu and Ma, Zejun and Zhang, Chao},
  title = {{SALMONN}: Towards Generic Hearing Abilities for Large Language Models},
  booktitle = {Proc. Int. Conf. Learn. Represent. (ICLR)},
  year = {2024},
  pages = {16607--16629},
  url = {https://openreview.net/forum?id=14rn7HpKVk}
}

@article{qwen25omni,
  author = {Xu, Jin and Guo, Zhifang and He, Jinzheng and Hu, Hangrui and He, Ting and Bai, Shuai and Chen, Keqin and Wang, Jialin and Fan, Yang and Dang, Kai and Zhang, Bin and Wang, Xiong and Chu, Yunfei and Lin, Junyang},
  title = {{Qwen2.5-Omni} Technical Report},
  journal = {arXiv preprint arXiv:2503.20215},
  year = {2025},
  doi = {10.48550/arXiv.2503.20215},
  url = {https://arxiv.org/abs/2503.20215}
}

@inproceedings{llaqo,
  author = {Zhang, Huan and Cheung, Vincent K. M. and Nishioka, Hayato and Dixon, Simon and Furuya, Shinichi},
  title = {{LLaQo}: Towards a Query-Based Coach in Expressive Music Performance Assessment},
  booktitle = {Proc. IEEE Int. Conf. Acoust., Speech Signal Process. (ICASSP)},
  year = {2025},
  pages = {1--5},
  publisher = {IEEE},
  doi = {10.1109/ICASSP49660.2025.10890522},
  url = {https://doi.org/10.1109/ICASSP49660.2025.10890522}
}

@inproceedings{vocalcoachbench,
  author = {Bang, Hayeon and Kim, Hounsu and Kim, Wonil and Nam, Juhan},
  title = {{VocalCoachBench}: Benchmarking Audio-Language Models on Expert Feedback for Singing},
  booktitle = {Advances in Neural Information Processing Systems ({NeurIPS}), Evaluations and Datasets Track},
  year = {2026},
  note = {To appear},
  url = {https://arxiv.org/abs/2609.04241}
}

@article{gupta2020leaderboard,
  author = {Gupta, Chitralekha and Li, Haizhou and Wang, Ye},
  title = {Automatic Leaderboard: Evaluation of Singing Quality Without a Standard Reference},
  journal = {IEEE/ACM Transactions on Audio, Speech, and Language Processing},
  year = {2020},
  volume = {28},
  pages = {13--26},
  doi = {10.1109/TASLP.2019.2947737},
  url = {https://doi.org/10.1109/TASLP.2019.2947737}
}

@inproceedings{vocalverse,
  author = {Wang, Zihao and Yuan, Ruibin and Geng, Ziqi and Li, Hengjia and Qu, Xingwei and Li, Xinyi and Chen, Songye and Fu, Haoying and Dannenberg, Roger B. and Zhang, Kejun},
  title = {Singing Timbre Popularity Assessment Based on Multimodal Large Foundation Model},
  booktitle = {Proc. ACM Int. Conf. Multimedia (ACM MM)},
  year = {2025},
  pages = {12227--12236},
  publisher = {Association for Computing Machinery},
  doi = {10.1145/3746027.3758148},
  url = {https://doi.org/10.1145/3746027.3758148}
}

@inproceedings{mullama,
  author = {Liu, Shansong and Hussain, Atin Sakkeer and Sun, Chenshuo and Shan, Ying},
  title = {Music Understanding {LLaMA}: Advancing Text-to-Music Generation with Question Answering and Captioning},
  booktitle = {Proc. IEEE Int. Conf. Acoust., Speech Signal Process. (ICASSP)},
  year = {2024},
  pages = {286--290},
  publisher = {IEEE},
  doi = {10.1109/ICASSP48485.2024.10447027},
  url = {https://doi.org/10.1109/ICASSP48485.2024.10447027}
}

@inproceedings{musicflamingo,
  author = {Ghosh, Sreyan and Goel, Arushi and Koroshinadze, Lasha and Lee, Sang-gil and Kong, Zhifeng and Santos, Joao Felipe and Duraiswami, Ramani and Manocha, Dinesh and Ping, Wei and Shoeybi, Mohammad and Catanzaro, Bryan},
  title = {{Music Flamingo}: Scaling Music Understanding in Audio Language Models},
  booktitle = {Proc. Int. Conf. Learn. Represent. (ICLR)},
  year = {2026},
  pages = {78958--78996},
  url = {https://proceedings.iclr.cc/paper_files/paper/2026/hash/7f64034009f4a5fa417a57e1a987c5cd-Abstract-Conference.html}
}

@article{funaudiochat,
  author = {{Tongyi Fun Team} and Chen, Qian and Cheng, Luyao and Deng, Chong and Li, Xiangang and Liu, Jiaqing and Tan, Chao-Hong and Wang, Wen and Xu, Junhao and Ye, Jieping and Zhang, Qinglin and Zhang, Qiquan and Zhou, Jingren},
  title = {{Fun-Audio-Chat} Technical Report},
  journal = {arXiv preprint arXiv:2512.20156},
  year = {2025},
  doi = {10.48550/arXiv.2512.20156},
  url = {https://arxiv.org/abs/2512.20156}
}

@inproceedings{melbandroformer,
  author = {Wang, Ju-Chiang and Lu, Wei-Tsung and Won, Minz},
  title = {{Mel-Band RoFormer} for Music Source Separation},
  booktitle = {Extended Abstracts for the Late-Breaking Demo Session of the 24th International Society for Music Information Retrieval Conference (ISMIR)},
  year = {2023},
  url = {https://ismir2023program.ismir.net/lbd_353.html}
}

@inproceedings{lora,
  author = {Hu, Edward J. and Shen, Yelong and Wallis, Phillip and Allen-Zhu, Zeyuan and Li, Yuanzhi and Wang, Shean and Wang, Lu and Chen, Weizhu},
  title = {{LoRA}: Low-Rank Adaptation of Large Language Models},
  booktitle = {Proc. Int. Conf. Learn. Represent. (ICLR)},
  year = {2022},
  url = {https://openreview.net/forum?id=nZeVKeeFYf9}
}

@misc{mossmusic,
  author = {{OpenMOSS Team}},
  title = {{MOSS-Music} Technical Report},
  year = {2026},
  howpublished = {\url{https://github.com/OpenMOSS/MOSS-Music}},
  note = {GitHub repository, accessed September 24, 2026}
}

@misc{mimoaudio,
      title={MiMo-Audio: Audio Language Models are Few-Shot Learners}, 
      author={LLM-Core-Team Xiaomi},
      year={2025},
      url={https://github.com/XiaomiMiMo/MiMo-Audio}, 
}

@misc{openai_models,
  author = {{OpenAI}},
  title = {{OpenAI API} Models},
  year = {2026},
  howpublished = {\url{https://developers.openai.com/api/docs/models}},
  note = {Accessed: September 24, 2026}
}

@misc{gptoss120b,
  author = {{OpenAI}},
  title = {{gpt-oss-120b \& gpt-oss-20b Model Card}},
  year = {2025},
  howpublished = {\url{https://openai.com/index/gpt-oss-model-card/}}
}

@inproceedings{adamw,
  author = {Loshchilov, Ilya and Hutter, Frank},
  title = {Decoupled Weight Decay Regularization},
  booktitle = {Proc. Int. Conf. Learn. Represent. (ICLR)},
  year = {2019},
  url = {https://openreview.net/forum?id=Bkg6RiCqY7}
}

@inproceedings{bleu,
  author = {Papineni, Kishore and Roukos, Salim and Ward, Todd and Zhu, Wei-Jing},
  title = {{Bleu}: a Method for Automatic Evaluation of Machine Translation},
  booktitle = {Proc. Annu. Meeting Assoc. Comput. Linguistics (ACL)},
  year = {2002},
  pages = {311--318},
  publisher = {Association for Computational Linguistics},
  doi = {10.3115/1073083.1073135},
  url = {https://aclanthology.org/P02-1040/}
}

@inproceedings{spice,
  author = {Anderson, Peter and Fernando, Basura and Johnson, Mark and Gould, Stephen},
  title = {{SPICE}: Semantic Propositional Image Caption Evaluation},
  booktitle = {Proc. Eur. Conf. Comput. Vis. (ECCV)},
  year = {2016},
  series = {Lecture Notes in Computer Science},
  volume = {9909},
  pages = {382--398},
  publisher = {Springer},
  doi = {10.1007/978-3-319-46454-1_24},
  url = {https://doi.org/10.1007/978-3-319-46454-1_24}
}

@inproceedings{selfbleu,
  author = {Zhu, Yaoming and Lu, Sidi and Zheng, Lei and Guo, Jiaxian and Zhang, Weinan and Wang, Jun and Yu, Yong},
  title = {{Texygen}: A Benchmarking Platform for Text Generation Models},
  booktitle = {Proc. Int. ACM SIGIR Conf. Res. Develop. Inf. Retr. (SIGIR)},
  year = {2018},
  pages = {1097--1100},
  publisher = {Association for Computing Machinery},
  doi = {10.1145/3209978.3210080},
  url = {https://doi.org/10.1145/3209978.3210080}
}

@inproceedings{yeh2000,
  author = {Yeh, Alexander},
  title = {More Accurate Tests for the Statistical Significance of Result Differences},
  booktitle = {Proc. Int. Conf. Comput. Linguistics (COLING)},
  year = {2000},
  volume = {2},
  doi = {10.3115/992730.992783},
  url = {https://aclanthology.org/C00-2137/}
}

@article{holm1979,
  author = {Holm, Sture},
  title = {A Simple Sequentially Rejective Multiple Test Procedure},
  journal = {Scandinavian Journal of Statistics},
  year = {1979},
  volume = {6},
  number = {2},
  pages = {65--70},
  url = {https://www.jstor.org/stable/4615733}
}

@inproceedings{jiang2023expert,
  author = {Jiang, Yucong},
  title = {Expert and Novice Evaluations of Piano Performances: Criteria for Computer-Aided Feedback},
  booktitle = {Proc. Int. Soc. Music Inf. Retr. Conf. (ISMIR)},
  year = {2023},
  pages = {367--374},
  publisher = {ISMIR},
  doi = {10.5281/zenodo.10265301},
  url = {https://archives.ismir.net/ismir2023/paper/000043.pdf}
}

@inproceedings{lerch2019music,
  author = {Lerch, Alexander and Arthur, Claire and Pati, Ashis and Gururani, Siddharth},
  title = {Music Performance Analysis: A Survey},
  booktitle = {Proc. Int. Soc. Music Inf. Retr. Conf. (ISMIR)},
  year = {2019},
  pages = {33--43},
  publisher = {ISMIR},
  doi = {10.5281/zenodo.3527735},
  url = {https://doi.org/10.5281/zenodo.3527735}
}

@inproceedings{llark,
  author = {Gardner, Joshua P. and Durand, Simon and Stoller, Daniel and Bittner, Rachel M.},
  title = {{LLark}: A Multimodal Instruction-Following Language Model for Music},
  booktitle = {Proc. Int. Conf. Mach. Learn. (ICML)},
  year = {2024},
  series = {Proceedings of Machine Learning Research},
  volume = {235},
  pages = {15037--15082},
  publisher = {PMLR},
  url = {https://proceedings.mlr.press/v235/gardner24a.html}
}

\end{document}